\documentclass[journal=jacsat,manuscript=article]{achemso}

\usepackage[version=3]{mhchem} 
\usepackage{booktabs}
\usepackage{multirow}
\usepackage{amsmath}
\usepackage{bm}
\usepackage{xcolor}

\author{Pritam Das}
\affiliation[IKST]
{Indo-Korea Science and Technology Center (IKST), Jakkur, Bengaluru 560065, India}
\author{Krishnamohan Thekkepat}
\affiliation[KIST]
{Electronic Materials Research Center, Korea Institute of Science and Technology, Seoul 02792, Republic of Korea}
\author{Young-Su Lee}
\affiliation[KIST2]{Energy Materials Research Center, Korea Institute of Science and Technology, Seoul 02792, Seoul, Republic of Korea.}
\author{Seung-Cheol Lee}
\email{leesc@kist.re.kr}
\affiliation[KIST]
{Electronic Materials Research Center, Korea Institute of Science and Technology, Seoul 02792, Republic of Korea}
\author{Satadeep Bhattacharjee}
\email{s.bhattacharjee@ikst.res.in}
\affiliation[IKST]
{Indo-Korea Science and Technology Center (IKST), Jakkur, Bengaluru 560065, India}

\title{Computational design of novel MAX phase alloys for potential hydrogen storage media combining first principles and cluster expansion methods}

\begin{document}

\begin{abstract}

Finding a suitable material for hydrogen storage at ambient atmospheric conditions is challenging for material scientists and chemists. In this work, using a first principles based cluster expansion approach, the  hydrogen storage capacity of Ti\textsubscript{2}AC (A = Al, Ti, Cr, Mn, Fe, Co, Ni, Cu, and Zn) MAX phase and its alloys were studied. We found that hydrogen is energetically stable in Ti-A layers in which the tetrahedral site consisting of one A atom and three Ti atoms is energetically more favorable for hydrogen adsorption than other sites in the Ti-A layer. Ti\textsubscript{2}CuC has the highest hydrogen adsorption energy than other Ti\textsubscript{2}AC phases. We find that 83.33\% Cu doped Ti\textsubscript{2}Al\textsubscript{x}Cu\textsubscript{1-x}C alloy structure is both energetically and dynamically stable and can store 3.66 wt\% hydrogen at ambient atmospheric conditions, which is higher than both Ti\textsubscript{2}AlC and Ti\textsubscript{2}CuC phase. These findings indicate that the hydrogen capacity of the MAX phase can be significantly improved by doping an appropriate atom species.

\end{abstract}


\section{Introduction}

Over the past decade, the MAX phase materials which are mainly composed of layered ternary carbides and nitrides are getting huge attention due to their unique properties that arise from the presence of both covalent and metallic bonds in the structure \cite{BARSOUM2000201, JEITSCHKO1964133}. The MAX phases (M\textsubscript{n+1}AX\textsubscript{n}, where M is early transition metals, A is a representative elements, X is C or N, and n = 1,2,3 $\cdots$) are ternary carbides and nitrides with a family of more than 80 members \cite{HADI2020109275}. MAX consists of early transition metals and group A elements, which are highly abundant and not hazardous to the environment. The wide availability and nontoxic nature of building elements of the MAX phase make the MAX phase cost-effective and environmentally friendly. These materials have a wide range of possible applications, including structural materials for high temperature applications, protective coatings and bond-coats for gas turbines, fuel cladding, solar receivers, electrical contacts, catalysts, and many more.
Depending on the different property between M-A (metallic) and M-X (covalent) bonding, MAX phases exhibit both ceramic and metallic property, including high electrical conductivity, oxidation resistance, catalysis, electronics, gas sensing, and energy storage \cite{doi:10.1021/acs.accounts.7b00481, doi:10.1021/acsenergylett.6b00247, doi:10.1021/acsnano.9b06394}. 
Some MAX phases, such as Ti\textsubscript{3}AlC\textsubscript{2}, Ti\textsubscript{2}AlC, Cr\textsubscript{2}AlC, also show exceedingly good oxidation resistance and oxidation-induced self-healing ability \cite{FARLE20171969}, and can be used as high-temperature structural materials. Moreover, MAX phase materials are expected to be applied to cladding and structural materials of the rapid reactor for molten lead cooling due to excellent resistance to neutron and He, Ar, Kr, I, Xe, Au plasma irradiation damage ability and excellent resistance to molten lead corrosion \cite{LIU201823254}. Also, some MAX phase films, for example, Ti\textsubscript{3}SiC\textsubscript{2} films, are able to be applied to the surface of metal bipolar plates to improve the corrosion resistance and conductivity of metal bipolar plates as well a considerable application prospect in commercial fuel cells\cite{HANSON2019195}. MAX phases contain two alternately stacked fundamental structural units: the non-stoichiometric transition metal carbide or nitride slabs in NaCl-type crystal structure and the close-packed A-group atomic plane \cite{BENTZEL20154107}. 

MAX phase has been mostly studied with metal hydride to enhance the hydrogen storage capacity or production. 7 wt\% Ti\textsubscript{3}AlC\textsubscript{2} MAX-phase composition with Mg significantly reduce dehydrogenation temperature of the composite \cite{LAKHNIK20227274}. Liu et al. show Ti\textsubscript{3}AlC\textsubscript{2}/Pd catalyst with 3 wt\% Pd loading had a much higher capacity for hydrogen production than conventional Pd nanoparticles \cite{nano12050843}. MAX phase behaves as a support structure for palladium nanoparticles for the hydrogen generation from alkaline formaldehyde solution at room temperature. The layered structures of the MAX phase can also be useful in the hydrogen storage field. Very few studies are available on the hydrogen storage performance of the MAX phase or MXene. Hydrofluoric acid incompletely etched multilayered MXene phase shows $\sim4$ wt\% hydrogen storage capacity at ambient condition and Ti\textsubscript{2}CT\textsubscript{x} (T is a functional group) can store 8.8 wt\% at room temperature and 60 bar H\textsubscript{2} \cite{Liu2021}. Single layer MXene can store up to 8.6 wt\% hydrogen by chemisorption of the H atom (1.7 wt \%), physisorption of the H\textsubscript{2} molecule (3.4 wt \%), and Kubas-type binding of the H\textsubscript{2} molecule (3.4 wt \%) \cite{doi:10.1021/jp409585v}. Ding et al. studied the hydrogen incorporation mechanism into the MAX phase Ti\textsubscript{3}AlC\textsubscript{2} by first-principles simulation and find out Ti-Al layers are suitable for hydrogen incorporation \cite{DING20166387}.

Among all phases, Ti\textsubscript{2}AlC is one of the most light-weight and oxidation-resistant layered ternary carbides and it has been extensively studied by many researchers \cite{WANG2021109788, https://doi.org/10.1111/jace.17582, BADIE2021117025}. Light-weight material can enhance hydrogen storage capacity in any structure. In this work, we investigate the hydrogen storage capacity of Ti\textsubscript{2}AlC in MAX phase and also possible doped 211 MAX structure with higher hydrogen storage capacity. We have employed first-principles simulation along with cluster expansion approach to find out possible doped MAX structures. We find that Hydrogen can be only incorporated in Ti-Al layers. Hydrogen in Ti-C layers is unstable and easily migrate to Ti-Al layers. 83.33\% doped TiAl\textsubscript{x}Cu\textsubscript{1-x}C structure has higher hydrogen storage capacity than Ti\textsubscript{2}AlC and Ti\textsubscript{2}CuC structure. In addition, the doped TiAl\textsubscript{x}Cu\textsubscript{1-x}C structure is very stable in nature and can be synthesized experimentally.

\section{Computational details}

\subsection{\textit{ab-initio} calculations}

All first-principles calculations have been performed using the density functional theory (DFT) as implemented in the Vienna Ab initio Software Package (VASP) \cite{Kresse1996CMS, Kresse1996PRB}. For the core and valence electrons, projected augmented wave (PAW) potentials and plane-wave basis sets were used with the Perdew–Burke–Ernzerhof (PBE) functional to get the electronic energy \cite{Kresse1999, Bl1994}. Corrections were made to account for long-range interactions using the semi-empirical Grimme D2 dispersion method and for non-spherical contributions to the PAW potentials that were built into the code \cite{Grimme2006}. All energies converged within a cutoff of 500 eV. The conjugate gradient algorithm is used for structural optimization \cite{Perdew1996}. The convergence criteria for energy and force are 10\textsuperscript{-7} eV and 0.01 eV\AA\textsuperscript{-1}, respectively. In all calculations, spin polarisation was enabled, with the only exception of the isolated H\textsubscript{2} closed shell molecule.

The adsorption energy (E\textsubscript{ads}) of hydrogen on the different MAX structure were calculated by: 

\begin{equation}
	E_{ads} =\frac{1}{n}\times\left(E_{MAX+nH}-\left( E_{MAX}+\frac{n}{2}E_{H_2}\right)\right)
	\label{eq:1}
\end{equation}

where E\textsubscript{MAX+nH} is the energy of the hydrogenated MAX structure, E\textsubscript{MAX} and E\textsubscript{H\textsubscript{2}} are the energies of the pristine structure and isolated H\textsubscript{2}, respectively, and n is the number of hydrogen atoms involved in the adsorption.

The atomic charges for all structures were calculated by Bader charge analysis, as suggested by Henkelman and co-workers \cite{Henkelman2006, Tang2009}. The formation energy of Ti\textsubscript{2}Al\textsubscript{x}Cu\textsubscript{1-x}C structure at different Cu concentration ere calculated by:

\begin{equation}
	E_{f} = E_{Ti\textsubscript{2}Al\textsubscript{x}Cu\textsubscript{1-x}C} - x E_{Ti\textsubscript{2}AlC} - (1-x) E_{Ti\textsubscript{2}CuC}
	\label{eq:2}
\end{equation}

where E is the energy of the pure or doped structure and x is the atomic fraction in the doped structure.

In order to estimate the release temperature, the chemical potential of the hydrogen in the solid phase and in the gas phase needs to be evaluated. The chemical potential energy of hydrogen as a function of temperature (T) and pressure (p) was obtained using standard statistical thermodynamics, via the formula \cite{Khan2016}:

\begin{equation}
	\mu(T,p)=\frac{1}{2}\left(E_{el}+E_{ZPE}+H^0(T)-H^0(0)-TS^0(T)+k_BTln\left(\frac{p}{p^0}\right)\right)
	\label{eq:u}
\end{equation}

where E\textsubscript{el} and E\textsubscript{ZPE} are the electronic and zero point energy of the hydrogen molecule as derived from DFT, H\textsuperscript{0} and S\textsuperscript{0} is  enthalpy and entropy at standard pressure p\textsuperscript{0} = 1 bar. The values of enthalpy and entropy at standard pressure can be found in JANAF thermochemical tables \cite{Chase1982}. The critical hydrogen chemical potential, $\mu^C_H$, at which the different MAX structure release hydrogen from the structure is given by $\mu^C_H=1/n(E_{MAX+nH}-E_{MAX})$ \cite{Williamson2004}.

\subsection{Calculation of formation energies of the solid solutions using the Cluster expansion approach}
A DFT-based technique cannot efficiently scan the energies of the immensely large configuration space that includes doped materials. A simple but efficient way is to construct an effective Hamiltonian using a relatively smaller database of DFT energies for these systems, which can then be used to scan a large configuration space. The cluster expansion method is an widely used method and can be used to calculate the energy of a large number of doped structures \cite{PhysRevB.49.8627, BADIE2021117025}.  The total energy of an alloy with configuration $\bm{\sigma}$ is calculated by constructing the so called cluster expansion effective Hamiltonian which is given by~\cite{CE1,CE2,CE3}

\begin{equation}
	E^{CE}(\bm{\sigma}) = J_{0} + \sum_{\alpha}J_{\alpha} \Biggl \langle \prod_{i \in \alpha} \sigma_{i} \Biggr \rangle=J_0+\sum_\alpha J_\alpha\Pi_\alpha
	\label{eq:3}
\end{equation}
Here J\textsubscript{$\alpha$} is known as the effective cluster interactions (ECI). J\textsubscript{0} represents ECI of an empty cluster. $\alpha$ can represent a single (point), pair or triplet cluster of sites. Each lattice site (i) is assigned an occupation variable $\sigma_{i}$ depending on whether it is occupied by an Al or Cu atom. $\Pi_\alpha$ are the cluster correlation functions.
To create a DFT database of Cu-doped Ti\textsubscript{2}AlC structures, the DFT energy of Cu doped 155 random structures of Ti\textsubscript{2}Al\textsubscript{x}Cu\textsubscript{1-x}C was calculated. Of these, 140 configurations were used to train the cluster expansion model and the remaining 15 were used to test the model. We performed 10-fold cross-validations to avoid over-fitting the data.
\par
To obtain the ECI's and the model we use the least absolute shrinkage and selection operator (LASSO) approach where the following minimization is performed to obtain the optimal $J_\alpha$,
\begin{equation}
{\bf J}_{opt}=\underset{{\bf J}}{\operatorname{arg min}} ||{\bf E_{DFT}}-\bf \Pi \bf J||^2_2+\lambda||J||_1
\label{lasso}
\end{equation}
Here $\lambda||{\bf J}||_1$ is the $l_1$ regularization term and $\lambda$ is an hyper-parameter. To obtain new stable Cu-doped materials we follow the following strategy: first we fit the DFT energies in a satisfactory way to the cluster expansion Hamiltonian given by Eq.\ref{eq:3}, next we generate a large pool of doped configurations (1000 ) and calculate the total energies of  such (Ti\textsubscript{2}Al\textsubscript{x}Cu\textsubscript{1-x}C) configurations using the fitted cluster expansion Hamiltonian. These energies  for different compositions are then further used to calculate the formation energies. The formation energy of all compositions was calculated using equation \ref{eq:2}.

\section{Results and Discussion}

\begin{figure}[htbp]
	\centering
	\includegraphics[width=1.0\linewidth]{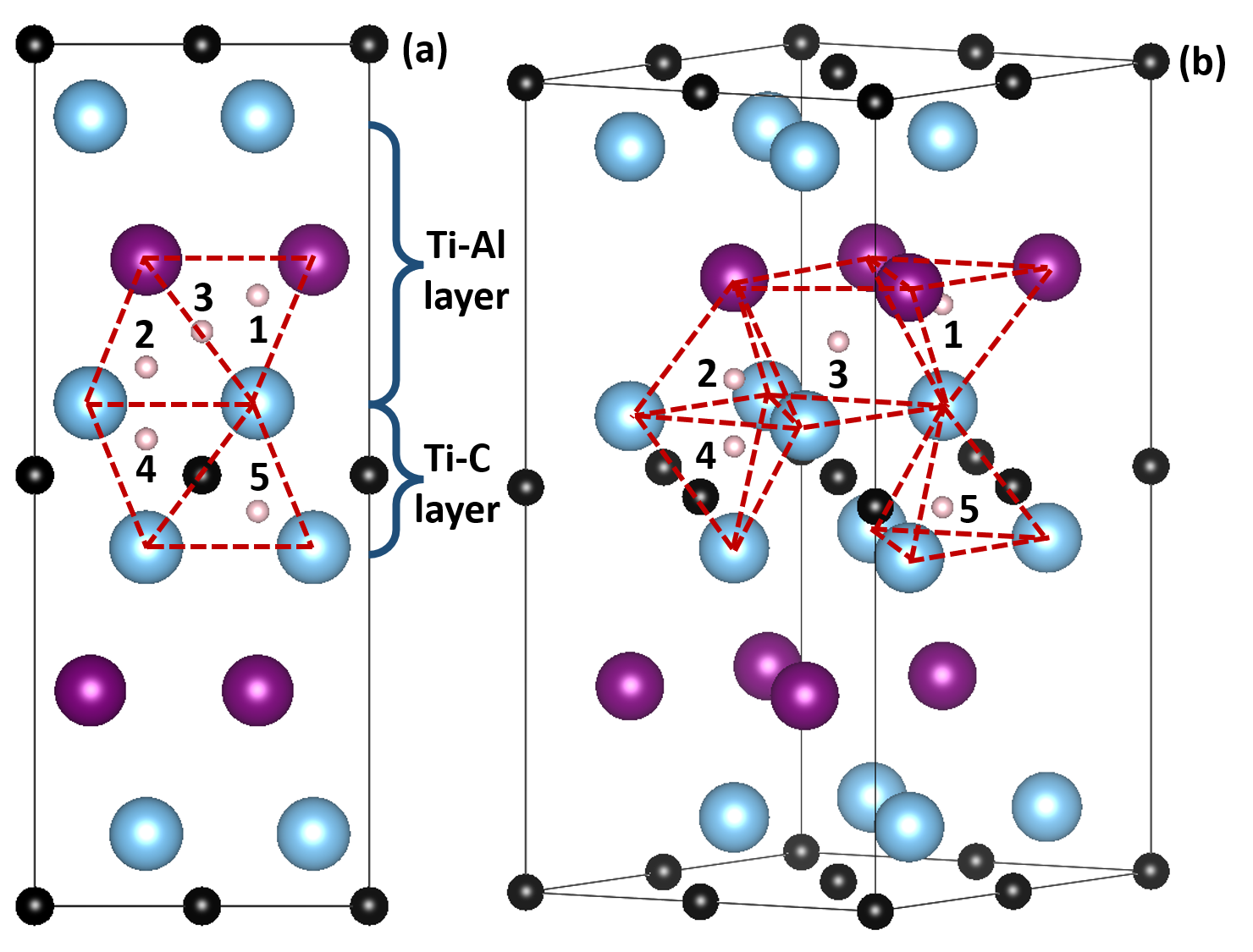}
	\caption{Crystal structure of 211 MAX phases and interstitial sites. Blue, purple, black, and white balls correspond to Ti, Al, C, and H, respectively.}
	\label{fig:ti2alc}
\end{figure}

In Ti\textsubscript{2}AlC has several interstitial sites and all of them fall into two parts: the interstitial sites in the Ti-Al and Ti-C layers. To find out possible hydrogen adsorption sites, all possible interstitial sites are considered. Figure \ref{fig:ti2alc} shows the possible H occupation sites in both Ti-Al and Ti-C layers. There are three kinds of interstitial sites in Ti-Al layers: the tetrahedral interstitial site I\textsubscript{tetr}-1 consists of three Al atoms and one Ti atom, the tetrahedral interstitial site I\textsubscript{tetr}-2 consists of three Ti atoms and one Al atom, and the octahedral interstitial site I\textsubscript{oct}-3 consists of three Al atoms and three Ti atoms. The Ti-C layers have two tetrahedral interstitial sites I\textsubscript{tetr}-4 and I\textsubscript{tetr}-5 consisting of three Ti atoms and one Ti atom from different Ti-C layers. The interstitial sites are designated as 1, 2, 3, 4, and 5, respectively in Fig. \ref{fig:ti2alc}.

The Hydrogen adsorption energy at different sites is calculated by Equation \ref{eq:1} and the results are shown in Table \ref{table:1}. Hydrogen atoms are exothermically adsorbed at all interstitial sites in Ti-Al layers. Among all interstitial sites in Ti-Al layers, I\textsubscript{tetr}-2 is more stable than I\textsubscript{oct}-3 and I\textsubscript{tetr}-1. In a relaxed structure, the H atom at I\textsubscript{tetr}-1 moves to the Al layers and locates in the center of the triangular interstitial site consisting of three atoms. In one Ti\textsubscript{2}AlC formula unit, a total of 6 interstitial sites are present in Ti-Al layers. Among 6 interstitial sites, two of each interstitial type (I\textsubscript{tetr}-1, I\textsubscript{tetr}-2, and I\textsubscript{oct}-3) are present. H atom in Ti-C layers is unstable and H atom move to I\textsubscript{tetr}-2 interstitial site after relaxation from I\textsubscript{tetr}-4 and I\textsubscript{tetr}-5.

\begin{table}[htbp]
	\caption{The adsorption energy of hydrogen at different interstitial sites of Ti\textsubscript{2}AlC}
	\centering
	\begin{tabular}{c c c c c c c} 
		\hline
		\textbf{Sites of hydrogen} & \multicolumn{3}{c}{\textbf{Ti-AL layer}} & & \multicolumn{2}{c}{\textbf{Ti-C layer}} \\
		\cmidrule{2-4}
		\cmidrule{6-7}
		& I\textsubscript{tetr}-1 & I\textsubscript{tetr}-2 & I\textsubscript{oct}-3 & & I\textsubscript{tetr}-4 & I\textsubscript{tetr}-5 \\
		\hline
		E\textsubscript{ads}(eV) & -0.06 & -0.40 & -0.23 & & unstable & unstable \\
		\hline
	\end{tabular}
	\label{table:1}
\end{table}

\begin{figure}[htbp]
	\centering
	\includegraphics[width=1.0\linewidth]{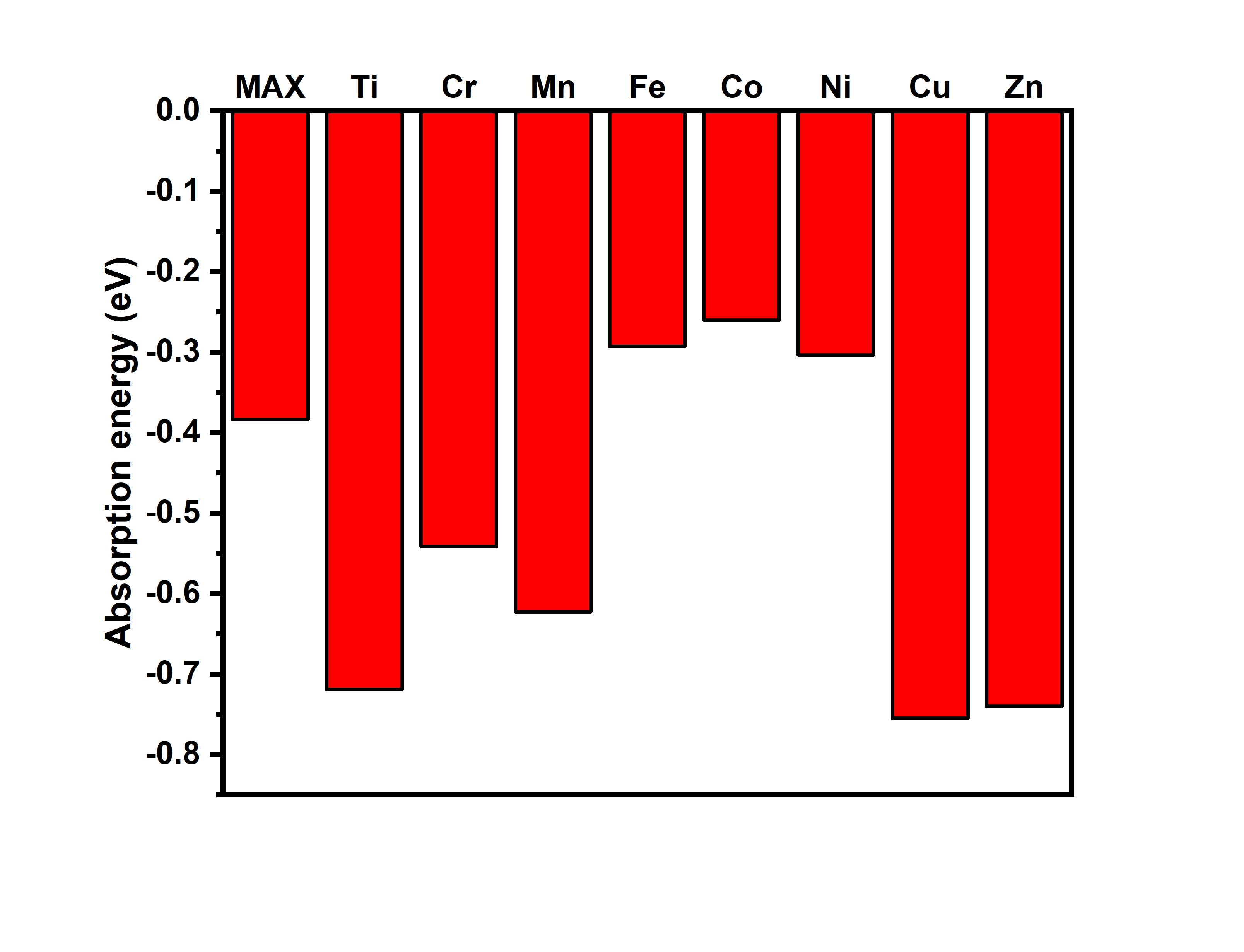}
	\caption{Hydrogen adsorption energy at I\textsubscript{tetr}-2 in different 211 MAX phases.}
	\label{fig:alloy}
\end{figure}

In the current study, we examined how the performance of hydrogen storage is affected when aluminium (Al) in the MAX phase is replaced with 3d-transition metals such as Ti, Cr, Mn, Fe, Co, Ni, Cu, and Zn atoms. Hydrogen adsorption energy was calculated by placing one hydrogen atom at I\textsubscript{tetr}-2 site in different Al substituted MAX phase. Figure \ref{fig:alloy} shows adsorption energy of different 211 MAX phase. Among all MAX phase Ti\textsubscript{2}CuC has the highest hydrogen adsorption energy of -0.76 eV/H. For hydrogen storage calculation, we compared hydrogen capacity between Ti\textsubscript{2}AlC and Ti\textsubscript{2}CuC phase.

\begin{figure}[htbp]
	\centering
	\includegraphics[width=1.0\linewidth]{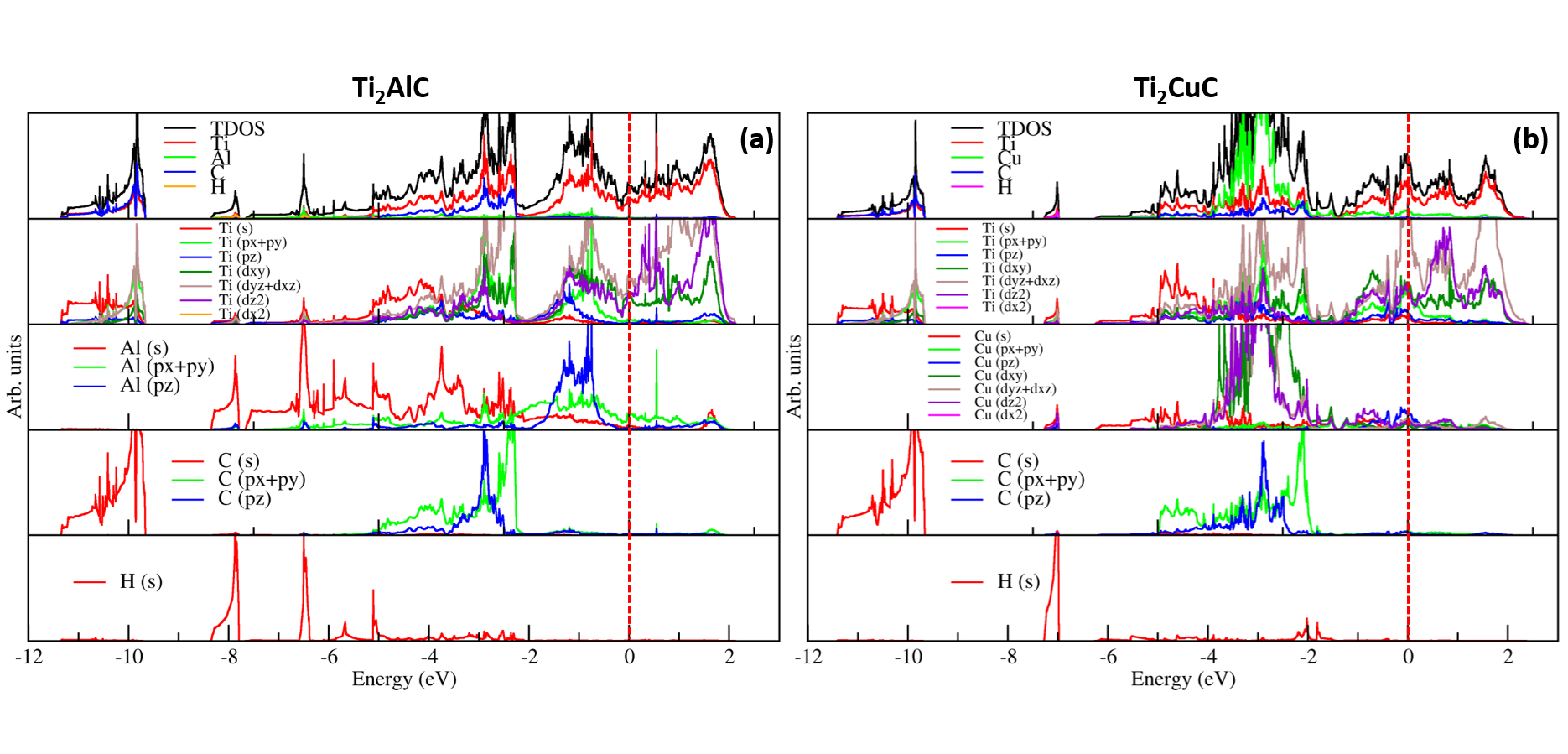}
	\caption{Electronic DOS of Ti\textsubscript{2}AlC (a) and Ti\textsubscript{2}CuC (b) structure.}
	\label{fig:dos}
\end{figure}

Next, we calculated electronic structure of Ti\textsubscript{2}AlC and Ti\textsubscript{2}CuC phase. As Ti has lower electronegativity and high oxidation states electron may transfer from Ti\textsubscript{2}C layer to Al or Cu atom. Figure \ref{fig:dos} shows total electronic density of states (DOS) and projected DOS of atomic orbitals of Ti\textsubscript{2}AlC and Ti\textsubscript{2}CuC MAX phases. In Ti\textsubscript{2}AlC phase, the Fermi level is contributed by not only Ti 3d but also Al 3s and Al 3p orbitals (fig. \ref{fig:dos} (a)). With increasing atomic number in the MAX phase, the contribution of Cu atom to the Fermi level decreases. The band center of s, p, and orbitals of Cu is located much deeper than Al (fig. \ref{fig:dos} (b)). The Ti 3d orbital of Ti\textsubscript{2}CuC phase contributed more to the Fermi level.

To find out the hydrogen adsorption mechanism, electron transfer between the hydrogen atom and MAX phase was calculated. The Bader charge analysis of Ti\textsubscript{2}AlC, Ti\textsubscript{2}CuC, and their hydrides a listed in Table \ref{table:2}. In the MAX phase, hydrogen is negatively charged because it accepts electron from neighboring atoms. The electron transfer to H is mostly contributed by Al and Cu atoms in Ti\textsubscript{2}AlC and Ti\textsubscript{2}CuC phase, respectively. In Ti\textsubscript{3}A tetrahedral, hydrogen present as H\textsuperscript{-} anion. Hydrogen in I\textsubscript{tetr}-2 site shows a greater stabilization effect through the Coulomb interaction. Hydrogen in Ti\textsubscript{2}CuC accepts lesser electrons than hydrogen in the Ti\textsubscript{2}AlC phase, which indicates the hydrogen bonding characteristic in the MAX phase changes from ionic to metallic. The bonding of hydrogen in the Ti\textsubscript{2}AlC phase is more ionic in nature, whereas in the Ti\textsubscript{2}CuC phase the bonging is metallic in nature.

\begin{table}[htbp]
	\caption{The Bader charge analysis of Ti\textsubscript{2}AlC, Ti\textsubscript{2}CuC, and their hydrides.}
	\centering
	\begin{tabular}{c c c c c c} 
		\hline
		 & & \textbf{Ti} & \textbf{Al} & \textbf{C} & \textbf{H}\\
		\hline
		\multirow{2}{4em}{Ti\textsubscript{2}AlC} & Ti\textsubscript{4}Al\textsubscript{2}C\textsubscript{2} & 2.03 & -1.21 & -2.84 & \\
		 & Ti\textsubscript{4}Al\textsubscript{2}C\textsubscript{2}H\textsubscript{1/2} & 2.03 & -1.06 & -2.82 & -1.34\\
		\hline
		 & & \textbf{Ti} & \textbf{Cu} & \textbf{C} & \textbf{H}\\
		\hline
		\multirow{2}{4em}{Ti\textsubscript{2}CuC} & Ti\textsubscript{4}Cu\textsubscript{2}C\textsubscript{2} & 2.03 & -1.31 & -2.75 & \\
		 & Ti\textsubscript{4}Cu\textsubscript{2}C\textsubscript{2}H\textsubscript{1/2} & 2.04 & -1.21 & -2.77 & -0.79\\
		\hline
	\end{tabular}
	\label{table:2}
\end{table}

\begin{figure}[htbp]
	\centering
	\includegraphics[width=1.0\linewidth]{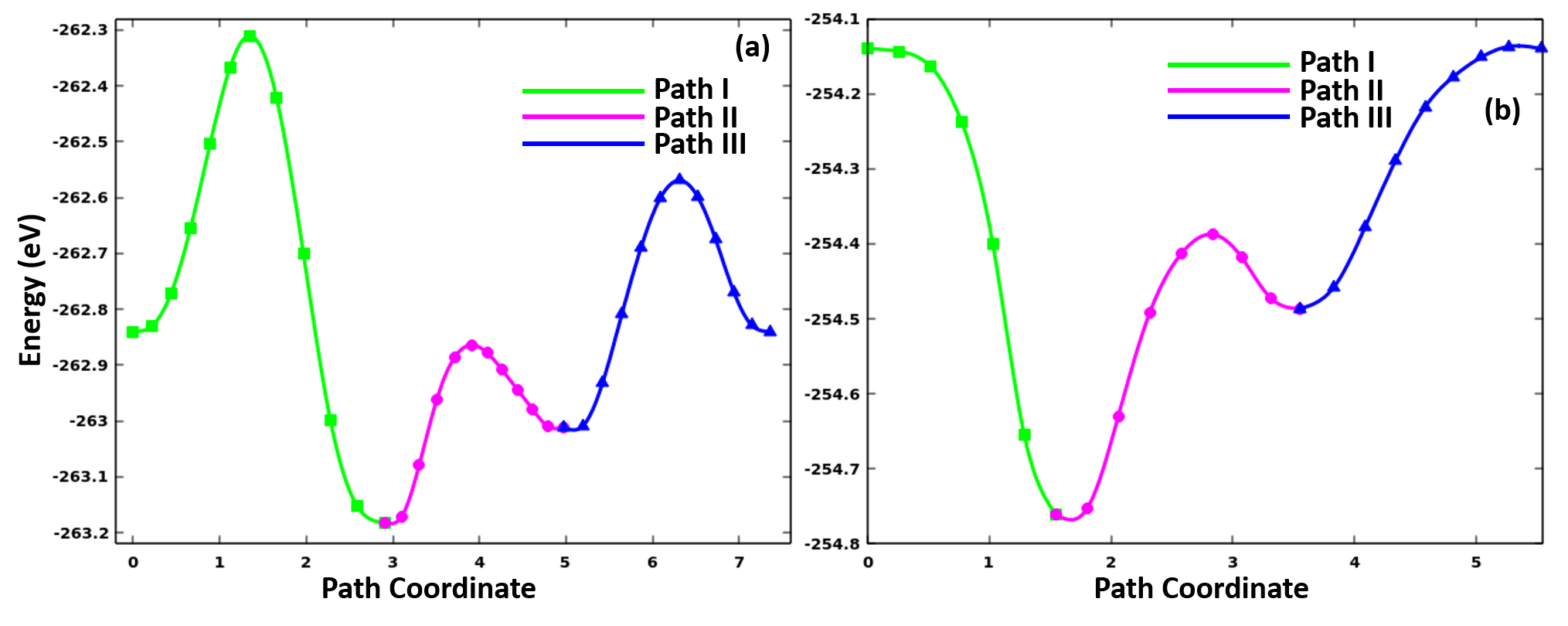}
	\caption{Diffusion of H in Ti-Al layers in Ti\textsubscript{2}AlC (a) and Ti\textsubscript{2}CuC (b) structure.}
	\label{fig:neb}
\end{figure}

To incorporate hydrogen in the material, the diffusion path of hydrogen in the material is necessary as the adsorption energy of the hydrogen in the materials. Ding et al. \cite{DING20166387} shows hydrogen diffusion in Ti-A layers of Ti\textsubscript{3}AlC\textsubscript{2} phase is following. Hydrogen firstly diffused from I\textsubscript{tetr}-1 to a nearest I\textsubscript{tetr}-2 (Path I), then diffused to a nearest I\textsubscript{oct}-3 (Path II), and finally to nearest I\textsubscript{tetr}-1 site (Path III). In this study same path is followed to find out the suitable MAX phase for easier diffusion of hydrogen among Ti\textsubscript{2}AlC and Ti\textsubscript{2}CuC. Figure \ref{fig:neb} (a) shows the H energy diffusion barrier for Ti\textsubscript{2}AlC. The energy diffusion barrier between I\textsubscript{tetr}-1 and I\textsubscript{tetr}-2 site is 0.53 eV. However, the opposite direction has a higher diffusion energy barrier of 0.87 eV. The diffusion energy barrier for I\textsubscript{tetr}-2 to I\textsubscript{oct}-3 is 0.32 eV, whereas the opposite path is easier with energy barrier of 0.15 eV. The path between I\textsubscript{oct}-3 and I\textsubscript{tetr}-1 is also easier than the path I. The energy barrier between I\textsubscript{oct}-3 and I\textsubscript{tetr}-1 is 0.44 eV and the opposite path is much easier with barrier energy 0.27 eV. Based on the energy barrier value, I\textsubscript{oct}-3 behaves as an intermediate position for hydrogen adsorption and desorption in Ti\textsubscript{2}AlC. The H atom of both I\textsubscript{tetr}-1 and I\textsubscript{tetr}-2 first diffuse to nearby I\textsubscript{oct}-3, then move to other I\textsubscript{tetr}-1 and I\textsubscript{tetr}-2 site. The activation energy of hydrogen diffusion in Ti-Al layers of Ti\textsubscript{2}AlC is much lower than $\alpha$-Ti and TiC \cite{papazoglou1968diffusion, hatano2006diffusion}, which could be useful for the hydrogen insertion mechanism in Ti\textsubscript{2}AlC.

Figure \ref{fig:neb} (b) shows the H energy diffusion barrier for Ti\textsubscript{2}CuC. The path between I\textsubscript{tetr}-1 and I\textsubscript{tetr}-2 site has very negligible energy barrier. However, the opposite direction has a higher diffusion energy barrier of 0.62 eV. The diffusion energy barrier for I\textsubscript{tetr}-2 to I\textsubscript{oct}-3 is 0.37 eV, whereas the opposite path is easier with energy barrier of 0.10 eV. The path between I\textsubscript{oct}-3 and I\textsubscript{tetr}-1 is also easier than the path I. The energy barrier between I\textsubscript{oct}-3 and I\textsubscript{tetr}-1 is 0.28 eV and the opposite path is much easier with barrier energy 0.003 eV. Similar to Ti\textsubscript{2}AlC structure in Ti\textsubscript{2}CuC I\textsubscript{oct}-3 behave as an intermediate position. But in the Ti\textsubscript{2}CuC structure, the H atom at I\textsubscript{tetr}-1 directly move to I\textsubscript{tetr}-2 (instead of I\textsubscript{oct}-3) since the energy barrier is negligible.

\begin{figure}[htbp]
	\centering
	\includegraphics[width=1.0\linewidth]{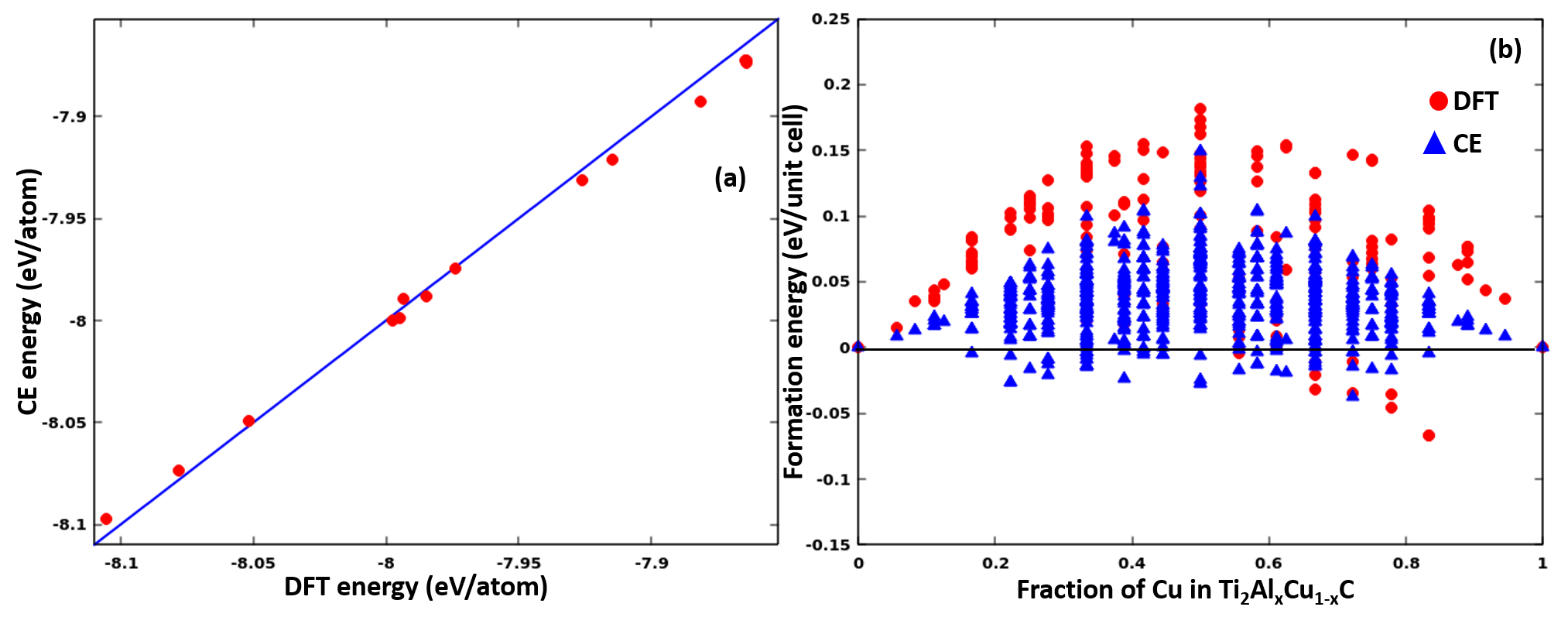}
	\caption{The agreement between the DFT recorded data and the cluster expansion (CE) model (a). The data points are clustered about the y = x line. Formation energy of Ti\textsubscript{2}Al\textsubscript{x}Cu\textsubscript{1-x}C at different Cu concentration (b). Both DFT (red) and CE (blue) formation energy are shown.}
	\label{fig:ce}
\end{figure}

\begin{figure}[htbp]
	\centering
	\includegraphics[width=1.0\linewidth]{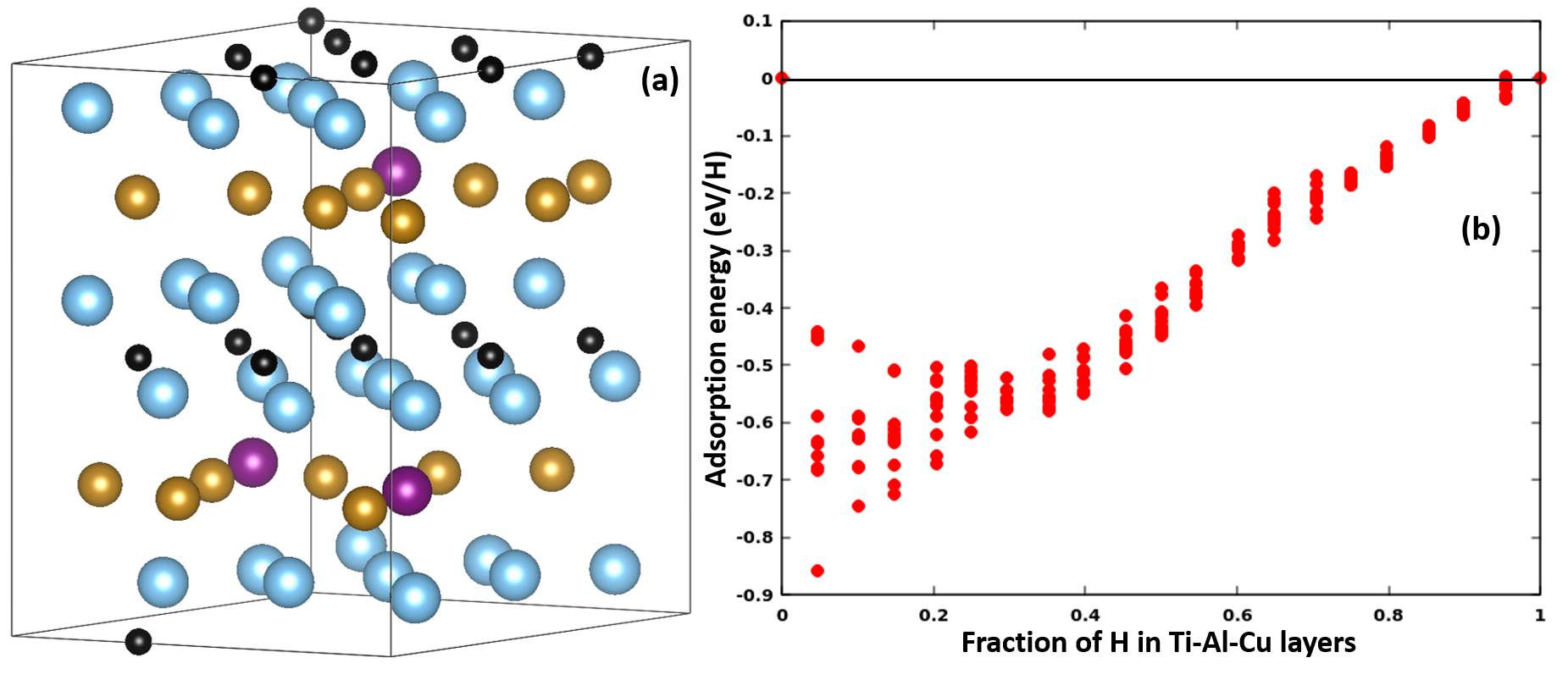}
	\caption{Crystal structure 83.33\% Cu doped optimized Ti\textsubscript{2}Al\textsubscript{0.17}Cu\textsubscript{0.83}C structure (a). Blue, purple, black, and brown balls correspond to Ti, Al, C, and Cu, respectively. Adsorption energy of hydrogen in Ti\textsubscript{2}Al\textsubscript{0.17}Cu\textsubscript{0.83}C at different H concentration (b).}
	\label{fig:h-alloy}
\end{figure}

The formation energy of Ti\textsubscript{2}CuC is higher than Ti\textsubscript{2}AlC. To find out the stable structures with copper substituted in Ti\textsubscript{2}AlC, cluster expansion (CE) Hamiltonian described in Eq.\ref{eq:3} was utilized to fit the DFT database of energies of Ti\textsubscript{2}Al\textsubscript{x}Cu\textsubscript{1-x}C alloys.  The accuracy of the CE model was verified by comparing it with DFT calculated energy of Cu doped Ti\textsubscript{2}AlC structure. The resulting CE energy is reasonably good at predicting DFT energy as the data points are clustered around the y=x line as shown in Fig. \ref{fig:ce} (a). The root-mean-square error (RMSE) of the fit was 0.0065 eV/atom. Using the fitted CE model, the formation energy of Cu doped Ti\textsubscript{2}AlC structures were calculated by increasing doped concentration by 5.56\%. Figure \ref{fig:ce} (b) shows the changes in formation energy in different Cu concentration obtained from both DFT calculation and CE. At 83.33\% doping concentration, the Ti\textsubscript{2}Al\textsubscript{x}Cu\textsubscript{1-x}C structure has the lowest formation energy, which is - 66.75 meV/unit cell (Fig. \ref{fig:h-alloy} (a)). The 83.33\% doped structure was relaxed by first-principles calculation before calculating hydrogen storage capacity and the dynamic stability of the structure was checked by phonon calculation. The 83.33\% Cu doped Ti\textsubscript{2}Al\textsubscript{x}Cu\textsubscript{1-x}C alloy structure is also dynamically stable in nature.

\begin{figure}[htbp]
	\centering
	\includegraphics[width=1.0\linewidth]{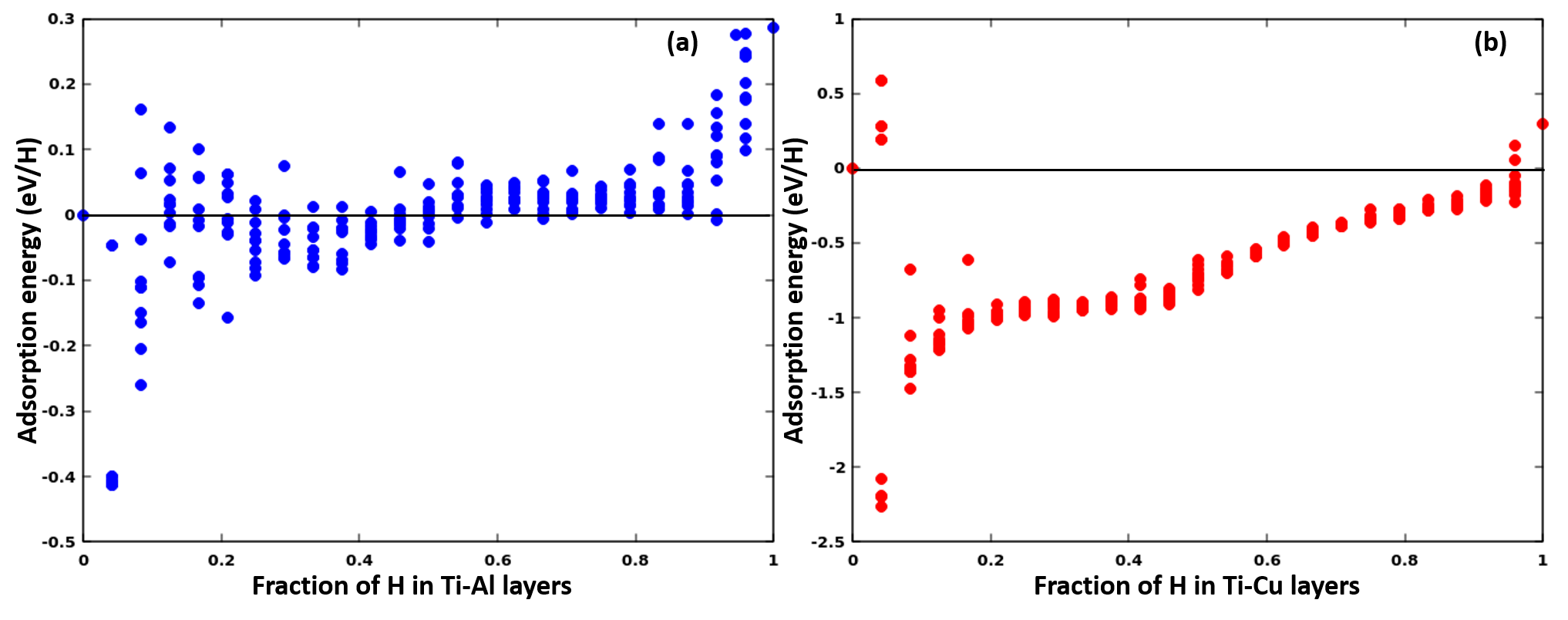}
	\caption{Adsorption energy of hydrogen in Ti\textsubscript{2}AlC (a) and Ti\textsubscript{2}CuC (b) at different H concentration.}
	\label{fig:h-pure}
\end{figure}

To find out the hydrogen storage capacity of Ti\textsubscript{2}AlC, Ti\textsubscript{2}CuC, and TiAl\textsubscript{x}Cu\textsubscript{1-x}C, all interstitial sites in the Ti-A layer are filled increasingly with the hydrogen atom and calculate the adsorption energy. Figure \ref{fig:h-pure} shows the changes in adsorption energy with increasing H concentration in Ti-A layer in Ti\textsubscript{2}AlC and Ti\textsubscript{2}CuC structure. In Fig. \ref{fig:h-pure}, all tetrahedral and octahedral sites in Ti-A layers filled with hydrogen atoms is designated as a 1.0 fraction of hydrogen H in Ti-A layers and a 0.0 fraction of hydrogen H in Ti-A layers implies no hydrogen atom present in the structure. The hydrogen adsorption energy was calculated by increasing the hydrogen number continuously in all structures and at which highest fraction of hydrogen the adsorption energy negative was considered as the maximum hydrogen capacity of the structure. The hydrogen weight percentage was calculated at that fraction of hydrogen for each structure. On the basis of hydrogen adsorption energy, Ti\textsubscript{2}AlC has a maximum hydrogen storage capacity of 2.62 wt\%, which is higher than previously reported chemisorbed H capacity in MAXene \cite{doi:10.1021/jp409585v}. In the case of Ti\textsubscript{2}CuC structure, hydrogen storage capacity in Ti-Cu increases by 3.38 wt\%. Figure \ref{fig:h-alloy} (b) shows the changes in adsorption energy with increasing H concentration in Ti-Al-Cu layer in 83.33\% Cu doped Ti\textsubscript{2}Al\textsubscript{0.17}Cu\textsubscript{0.83}C structure. The Ti\textsubscript{2}Al\textsubscript{0.17}Cu\textsubscript{0.83}C has highest hydrogen storage capacity of 3.66 wt\% and Ti\textsubscript{2}Al\textsubscript{0.17}Cu\textsubscript{0.83}CH\textsubscript{5.72} phase. In the Ti\textsubscript{2}Al\textsubscript{0.17}Cu\textsubscript{0.83}CH\textsubscript{5.72} phase, all I\textsubscript{tetr}-2 sites are filled by hydrogen atoms and 94.44\% and 91.67\% of I\textsubscript{oct}-3 and I\textsubscript{tetr}-1 sites are occupied by hydrogen atoms, respectively.

\begin{figure}[htbp]
	\centering
	\includegraphics[width=1.0\linewidth]{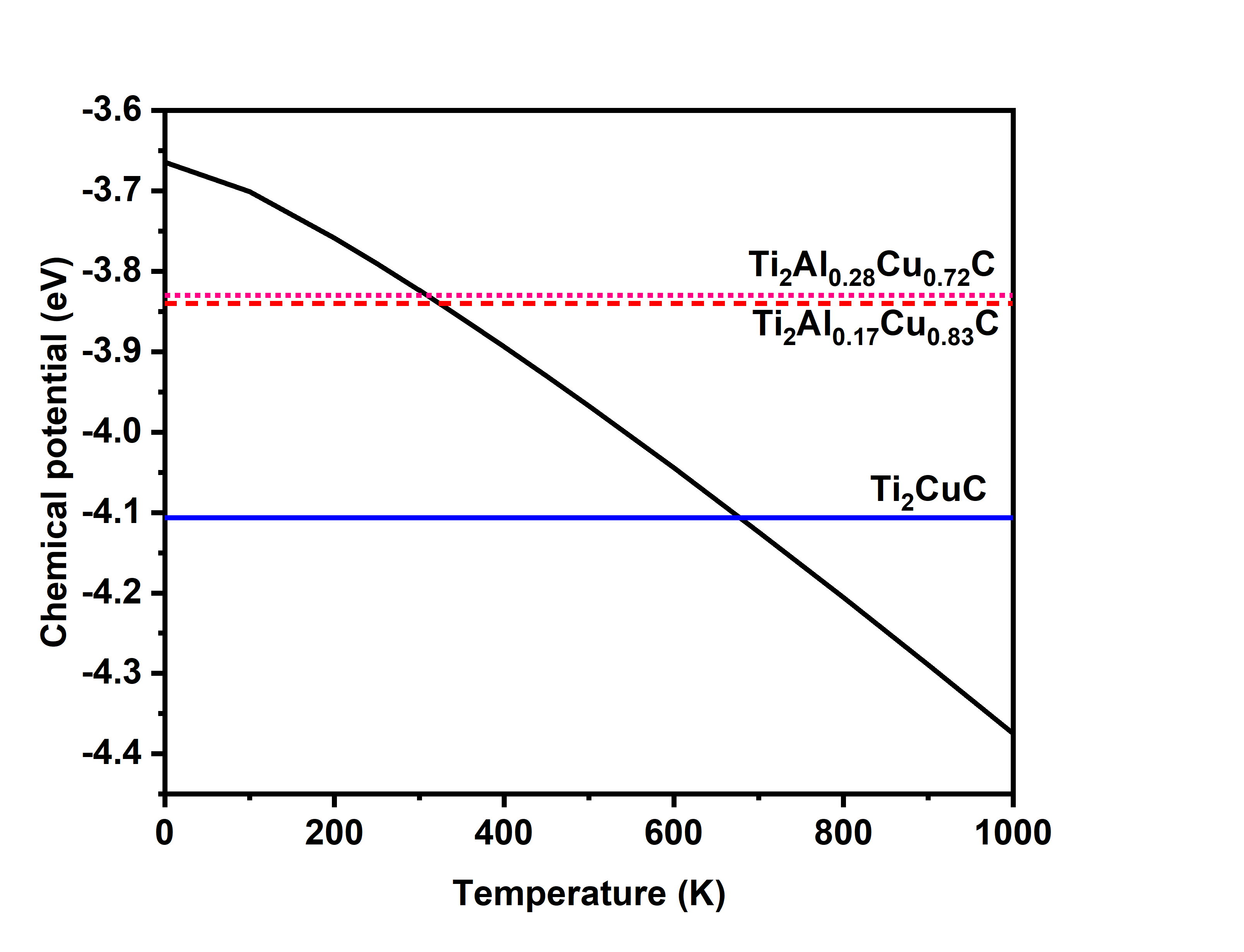}
	\caption{The temperature dependence of the chemical potential of hydrogen (black line). Horizontal lines marks the value of $\mu^C_H$ different MAX phase.}
	\label{fig:rt}
\end{figure}

The thick black curve in Fig \ref{fig:rt} shows the variation of the chemical potential of hydrogen atoms with the temperature at 1 bar pressure. The value of $\mu^c_H$ for different MAX phases at 50\% storage capacity is plotted by horizontal lines. The intersection between the these lines and the hydrogen chemical potential curve indicates the corresponding critical temperature of hydrogen release from the particular surface at 1 bar pressure \cite{Williamson2004}. At 50\% storage capacity, Ti\textsubscript{2}AlC phase has a very low critical hydrogen chemical potential (-3.41 eV). Whereas, the hydrogen release temperature of Ti\textsubscript{2}CuC phase was 677 K, which is quite higher than the operating temperature of the hydrogen fuel cell \cite{Eberle2012}. The Ti\textsubscript{2}Al\textsubscript{0.17}Cu\textsubscript{0.83}C alloy structure has a 326 K hydrogen release temperature, which is ideal for hydrogen storage applications. The hydrogen release temperature of  the lowest energy structure, Ti\textsubscript{2}Al\textsubscript{0.28}Cu\textsubscript{0.72}C obtained  using CE method was also calculated. The Ti\textsubscript{2}Al\textsubscript{0.28}Cu\textsubscript{0.72}C structure has a hydrogen release temperature of 311K, which is lower than Ti\textsubscript{2}Al\textsubscript{0.17}Cu\textsubscript{0.83}C structure.

\section{Conclusions}
In summary, we performed first-principles based cluster expansion calculation to study the hydrogen storage capacity in the MAX phase. The hydrogen atom is thermodynamically  stable in Ti-A layers of Ti\textsubscript{2}AC phase. Hydrogen atoms located in the Ti-C layer moved to the Ti-A layers. Among three interstitial sites in Ti-A layers, I\textsubscript{tetr}-2 site is most favorable for hydrogen storage. DFT calculation shows among different Ti\textsubscript{2}AC phases. Ti\textsubscript{2}CuC has the highest hydrogen adsorption energy due to the deeper p-orbital position of the Cu atom. Also, this phase is more preferable for the hydrogen release in comparison to Ti\textsubscript{2}AlC due to lower Coulomb interaction. The diffusion of hydrogen in Ti\textsubscript{2}CuC phase is more easier than Ti\textsubscript{2}AlC due to the lower energy barrier. The most stable Cu doped Ti\textsubscript{2}AlC alloy has been found by the cluster expansion model. The 83.33\% Cu doped Ti\textsubscript{2}Al\textsubscript{x}Cu\textsubscript{1-x}C alloy structure is both energetically and dynamically stable and can store 3.66 wt\% hydrogen at ambient atmospheric conditions, which is higher than both Ti\textsubscript{2}AlC and Ti\textsubscript{2}CuC phase. The 83.33\% Cu doped Ti\textsubscript{2}Al\textsubscript{x}Cu\textsubscript{1-x}C alloy structure is a promising 211 MAX phase for hydrogen storage application.

\section{Acknowledgement}
 This work was supported by the Korea Institute of Science and Technology (Grant number 2E31851), GKP (Global Knowledge Platform, Grant number 2V6760) project of the Ministry of Science, ICT and Future Planning.
\bibliography{achemso}

\end{document}